\documentclass[a4paper,aps,prl,reprint,superscriptaddress,showpacs]{revtex4-1}
\usepackage{amsmath}
\usepackage{bm}
\usepackage{multirow}
\usepackage{braket}
\usepackage{graphicx}
\usepackage{booktabs}
\usepackage{epstopdf}

\usepackage[unicode=true,colorlinks=true,linkcolor=blue,citecolor=blue,urlcolor=black]{hyperref}
\usepackage{newtxtext}
\usepackage{newtxmath}

\begin{document}
\title{Quartic Anharmonicity of Rattlers and Its Effect on Lattice Thermal Conductivity of Clathrates from First Principles}
\author{Terumasa Tadano}
\email{TADANO.Terumasa@nims.go.jp}
\affiliation{International Center for Young Scientists (ICYS), National Institute for Materials Science, Tsukuba 305-0047, Japan}
\affiliation{Research and Services Division of Materials Data and Integrated System (MaDIS), National Institute for Materials Science, Tsukuba 305-0047, Japan}

\author{Shinji Tsuneyuki}
\affiliation{Department of Physics, The University of Tokyo, Tokyo 113-0033, Japan}
\affiliation{Institute for Solid State Physics, The University of Tokyo, Kashiwa 277-8581, Japan}
\begin{abstract}
We investigate the role of the quartic anharmonicity in lattice dynamics and thermal transport of type-I clathrate Ba$_{8}$Ga$_{16}$Ge$_{30}$
based on  \textit{ab initio} self-consistent phonon calculations. 
We show that the strong quartic anharmonicity of rattling guest atoms causes the hardening of vibrational frequencies of
low-lying optical modes and thereby affects calculated lattice thermal conductivities $\kappa_{L}$ significantly, resulting in
an improved agreement with experimental results including the deviation from $\kappa_{L}\propto T^{-1}$ at high temperature.
Moreover, our static simulations with various different cell volumes shows a transition from crystal-like to \textit{glasslike} $\kappa_{L}$
around 20 K. Our analyses suggest that the resonance dip of $\kappa_{L}$ observed in clathrates with large guest-free-spaces is attributed mainly
to the strong three-phonon scattering of acoustic modes along with the presence of higher-frequency dispersive optical modes.
\end{abstract}
\maketitle


Intermetallic clathrates are a class of materials possessing a cage-like structure incorporating 
a guest atom, which is loosely bound inside an oversized cage and makes a large amplitude thermal vibration called ``rattling''. 
Experimental and theoretical studies have evidenced that 
rattling guest atoms cause characteristic lattice dynamics of clathrates including 
low-frequency vibrational modes showing significant temperature ($T$) dependence~\cite{Takasu2006,Christensen2008,Takabatake2014,2016PhRvB..93i4303W} and 
very low lattice thermal conductivity (LTC, $\kappa_{L}$)~\cite{Dong2001,Takabatake2014,PhysRevLett.114.095501}.
LTCs of clathrates are not only very low but also show exceptional and diverse $T$-dependence.
For example, $\kappa_{L}$ of electron-doped type-I clathrate Ba$_{8}$Ga$_{16}$Ge$_{30}$ (BGG)
shows a peak around 20 K followed by a decreasing region at higher temperature~\cite{2001PhRvB..63x5113S,Avila2006},
just like many crystalline solids. However, in the temperature region above $\sim$100 K, 
the $T$-dependence is much milder than  $\kappa_{L} \propto T^{-1}$ of
typical crystalline materials~\cite{PhysRevB.80.125205}.
More exceptional $T$-dependence has been reported for type-I clathrates X$_{8}$Ga$_{16}$Ge$_{30}$ (X=Sr, Eu)~\cite{1998ApPhL..73..178N,2001PhRvB..63x5113S} and Ba$_{8}$Ga$_{16}$Sn$_{30}$~\cite{2008ApPhL..92d1901A}.
In these materials, the LTCs behave like a typical glass, showing 
a plateau region or a ``resonance dip''~\cite{1998ApPhL..73..178N} near $\sim$20 K, 
an increasing trend in $\sim$20--100 K, and a nearly $T$-independent region above 100 K.
These unconventional and diverse thermal transport in intermetallic clathrates has been attributed to the difference in
the guest-free-space of the rattling atoms~\cite{2001PhRvB..63x5113S,2007PhRvB..75s5210S,2010PhRvB..81t5207S}, 
the host-guest coupling strength~\cite{2004PhRvB..70n0201B,PhysRevB.96.064306}, 
and the magnitude of static/dynamical disorders~\cite{2004PhRvB..70n0201B,Christensen:2016dv}.
Despite these continuous efforts, the origin of the unusual LTCs of clathrates still remains unclear.

Recently, \textit{ab initio} calculation of LTC based on the Peierls-Boltzmann theory~\cite{Peierls} 
has established itself as a convenient way to predict/analyze 
thermal transport phenomena in solids~\cite{Lindsay:2016iv}.
Although the validity of the Boltzmann theory is limited to the cases where the phonon quasiparticle picture is well established~\cite{Allen:1993gt},
it has reproduced experimental LTC of BGG in a relatively low temperature region~\cite{PhysRevLett.114.095501}. 
However, the \textit{ab initio} Boltzmann approach considerably underestimated $\kappa_{L}$ above $\sim 100$ K.
A similar underestimation has also been reported in a more recent study on another type-I clathrate Ba$_{7.81}$Ge$_{40.67}$Au$_{5.33}$~\cite{Lory:2017eh}. 
These results clearly indicate the necessity of an improved theoretical approach.
One of the most problematic approximations made in the conventional Boltzmann approach is the omission of the quartic anharmonicity.
Indeed, the atomic displacement factor of guest atoms is so significant that the quartic anharmonicity cannot be neglected anymore in clathrate.

Here, we report volume- and temperature-dependence of $\kappa_{L}$ of type-I clathrate BGG obtained from first-principles calculation,
where the temperature renormalization of the vibrational frequency by the quartic anharmonicity is considered using
the self-consistent phonon (SCP) theory~\cite{Hooton,1966PhRvL..17...89K,PhysRevLett.100.095901,Errea:2014ke,2015PhRvB..92e4301T,vanRoekeghem:2016kd}.
We show that the strong quartic anharmonicity of rattling motions makes the deviation from $\kappa_{L} \propto T^{-1}$ at high temperature.
Our computational analyses also indicate that the resonance dip observed in clathrates with large cage-sizes can be attributed to 
the strong three-phonon scattering of acoustic modes along with the presense of high-frequency dispersive optical modes and
the phonon-boundary scatterings.

\begin{figure*}
\centering
\includegraphics[width=0.80\textwidth,clip]{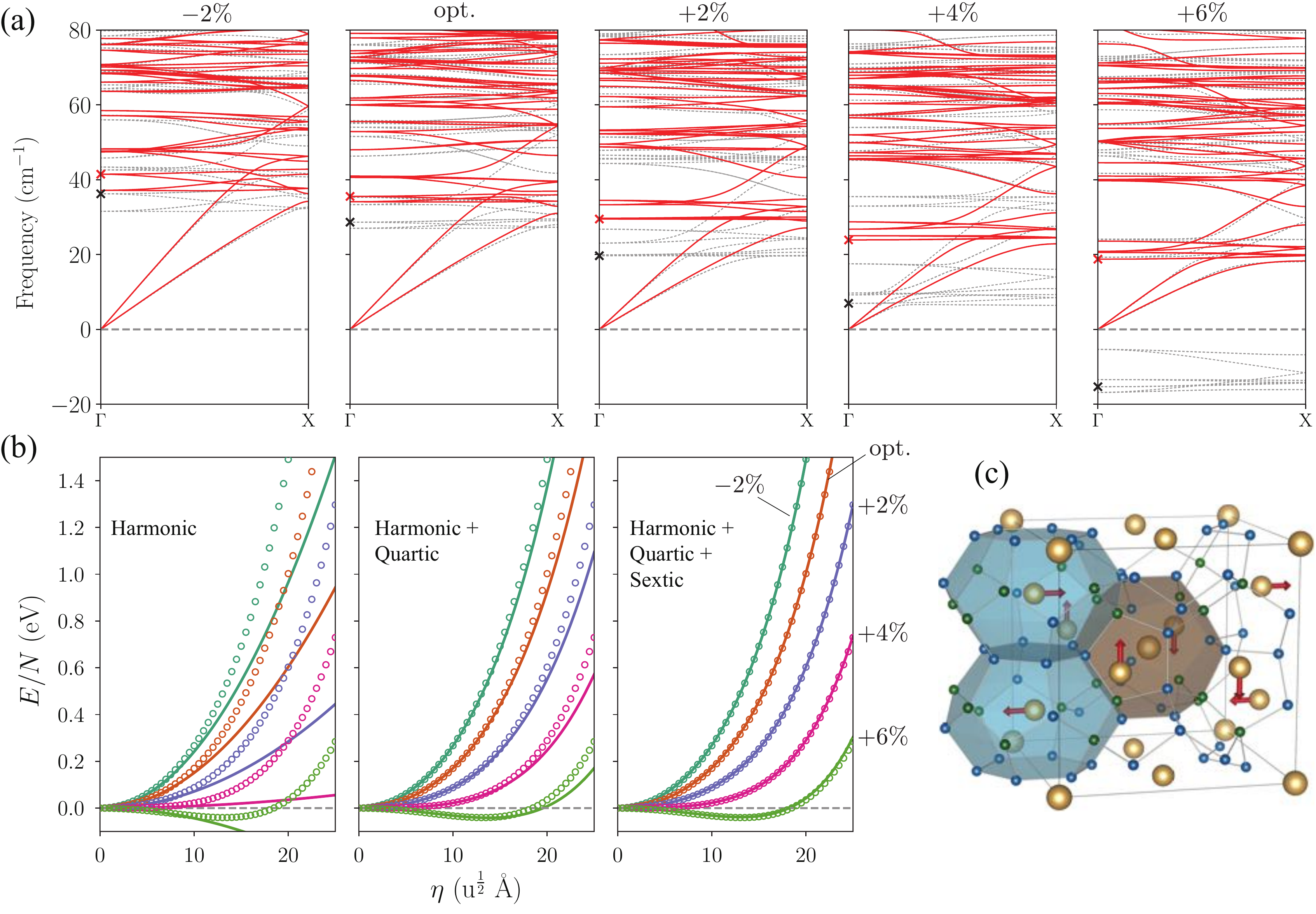}
\caption{(Color online) Lattice dynamics of type-I clathrate Ba$_{8}$Ga$_{16}$Ge$_{30}$ with various lattice constants. (a) 
Phonon dispersion curves in the low-frequency region along the $\Gamma$--X line.
The dotted lines show the harmonic phonon frequencies and the solid lines are the result of the SCP calculations at 300 K.
The cross symbols ``$\times$'' indicate the positions of the lowest Raman active $T_{2g}$ modes. 
(b) Potential energy surface of the lowest $T_{2g}$ mode. The circles ``$\circ$'' represent the results of DFT calculations and
the lines are obtained from the harmonic and anharmonic force constants.
(c) Crystal structure of Ba$_{8}$Ga$_{16}$Ge$_{30}$ (with \textsc{VESTA}~\cite{Momma:db5098}). 
The red arrows indicate the displacement pattern of the $T_{2g}$ mode.}
\label{fig:dispersion}
\end{figure*}


We start by giving a brief overview of the SCP method.
Following our recent implementation~\cite{2015PhRvB..92e4301T},
we calculate the $T$-dependent \textit{anharmonic} frequencies $\Omega_{q}(T)$ and polarization vectors $\bm{\epsilon}_{q}(T)$ by diagonalizing the matrix $\bm{V}_{\bm{q}}$ defined as
\begin{equation}
V_{\bm{q}jj'}= \omega^{2}_{\bm{q}j}\delta_{jj'}+\frac{1}{2}\sum_{q_{1}}\Phi (\bm{q}j;-\bm{q}j';q_{1};-q_{1})\braket{Q_{q_1}^{*}Q_{q_1}}. \label{eq:SCP1}
\end{equation}
Here, $\omega_{\bm{q}j}$ is the harmonic phonon frequency with the crystal momentum $\bm{q}$ and the branch index $j$, 
and $q$ is the shorthand notation for $(\bm{q},j)$ satisfying $q = (\bm{q}, j)$ and $-q=(-\bm{q},j)$. 
$\Phi(\bm{q}j;-\bm{q}j';q_{1};-q_{1})$ is the reciprocal representation of the fourth-order interatomic force constants (IFCs).
$\braket{Q_{q}^{*}Q_{q}}=\frac{\hbar}{2\Omega_{q}}[1+2n(\Omega_{q})]$ is the mean square displacement of the normal coordinate $Q_q$,
where $n(\omega) = 1 / (e^{\beta\hbar\omega}-1)$ is the Bose-Einstein distribution function, $\beta=1/kT$ with the Boltzmann constant $k$, and $\hbar$ is the reduced Planck constant, respectively.
The frequency renormalization by the SCP theory includes the effect of an infinite set of Feynman diagrams generated from the loop diagram,
which is a first-order correction by the quartic anharmonicity~\cite{Tadano:2018ex}. Therefore, it should be more important than the four-phonon scattering, which is a second-order correction, especially for low-$\kappa_{L}$ materials.
More details about the present SCP formalism are described elsewhere~\cite{2015PhRvB..92e4301T,Tadano:2018ex}.


All of the DFT calculations were conducted using the \textsc{vasp} code~\cite{Kresse1996}, with the GGA-PBE functional~\cite{PBE1996} and the projector augmented wave (PAW) method~\cite{PAW1994,Kresse1999}, and the phonon calculations were performed using the \textsc{alamode} package~\cite{alamode,Tadano2014}.
An ordered unit cell containing 54 atoms (space group $Pm\bar{3}n$) was employed for modeling lattice anharmonicity and phonon thermal transport. To investigate the effect of the guest free space on phonon properties,
the lattice constant of BGG was compressed/expanded from the optimized value (10.954 \AA, hereafter called ``opt.'') by $-2$\% to $+6$\% in steps of 2\%.
Harmonic and anharmonic IFCs were estimated using the compressive sensing lattice dynamics method~\cite{PhysRevLett.113.185501}.
More detailed computational procedures are provided as the Supplemental Information (SI)~\cite{supplement}.


Calculated harmonic phonon dispersion curves of BGG in the low-frequency region ($\leq 80$ cm$^{-1}$) are shown in Fig.~\ref{fig:dispersion}(a) by dotted lines.
With increasing the lattice constant, the frequencies of the low-lying optical modes associated with rattlers decrease.
In the largest-volume case, twelve phonon modes become unstable $(\omega_{q}^{2} < 0)$, all of which can be well characterized as collective motions of Ba(2) atoms inside the tetrakaidecahedral cages (see Fig.~\ref{fig:dispersion}(c)).
To see the volume dependence of the lattice anharmonicity more directly, 
we also calculated the potential energy surface (PES) of the Raman active $T_{2g}$ guest mode by displacing atoms
by $\bm{u}_{\kappa} = M_{\kappa}^{-1/2}\bm{e}_{q,\kappa}\eta$ with $\eta$ being the amplitude of the normal coordinate of the $T_{2g}$ mode. 
In Fig.~\ref{fig:dispersion}(b), we compare the PESs calculated by DFT and those calculated from the IFCs.
The left panel of Fig.~\ref{fig:dispersion}(b) shows that
the harmonic approximation fails to capture the actual shape of the PES, indicating the significance of the anharmonicity.
Indeed, if we include the contribution from the fourth-order IFCs, 
we obtain overall good agreements with the DFT results (middle panel).
Moreover, the sixth-order IFCs further improve the accuracy of the Taylor expansion potential (right panel).
Since the correction by the sextic terms is minor, however, 
we only considered the dominant quartic terms in the SCP calculations.

We calculated finite-temperature phonon dispersion curves by solving the SCP equation [Eq.~(\ref{eq:SCP1})] at various temperatures.
The SCP dispersion curves at 300 K are shown in Fig.\ref{fig:dispersion}(a) by solid lines.
The quartic anharmonicity generally increases the frequencies of the low-lying rattling modes,
which can be attributed to the dominant and positive contribution from the diagonal term of the quartic coefficient;
$\Phi(\bm{0}j;\bm{0}j;\bm{0}j;\bm{0}j) > 0$ for the low-lying optical modes $j$. 
With increasing the cage size, the quartic component of the PES becomes more important as shown in Fig.~\ref{fig:dispersion}(b), leading to the greater frequency shifts.
For the higher-frequency modes above 80 cm$^{-1}$, the anharmonic renormalization was turned out to be relatively small (see the SI~\cite{supplement}).
The present result is the first realization of the \textit{ab initio} SCP calculation for a complex host-guest structure.

According to the Raman study of Takasu \textit{et al.}~\cite{Takasu2006}, 
the frequency of the $T_{2g}$ guest mode in BGG increases from 31 cm$^{-1}$ at 2 K to 34 cm$^{-1}$ at room temperature.
The SCP theory for the ``opt.'' case gives 29.8 cm$^{-1}$ at 0 K and 35.5 cm$^{-1}$ at 300 K,
which agree well with the experimental values especially given that the present SCP theory 
neglects the intrinsic frequency shift by the cubic anharmonicity and the quasiharmonic effect.
To investigate the significance of the intrinsic frequency shift by the cubic terms,
we also calculated the lowest-order correction by the cubic anharmonicity from the real part of the bubble self-energy as 
$\Delta_{q} = -\mathrm{Re}\Sigma_{q}^{(\mathrm{B})}(\Omega_{q})$ with 
$\Sigma_{q}^{(\mathrm{B})}(\omega)$ being defined as 
\begin{align}
\Sigma_{q}^{(\mathrm{B})}(\omega) &= \frac{\hbar}{2N} \sideset{}{'}\sum_{q_{1},q_{2},s=\pm1} \frac{|\Phi(-q;q_{1};q_{2})|^{2}}{8\Omega_{q}\Omega_{q_{1}}\Omega_{q_{2}}}\notag \\
&\hspace{2mm} \times \left[ \frac{n_{1} + n_{2} + 1}{s\omega_{c} + \Omega_{q_{1}} + \Omega_{q_{2}}} 
 - \frac{n_{1} - n_{2}}{s\omega_{c} + \Omega_{q_{1}} - \Omega_{q_{2}}} \right].
\label{eq:bubble}
\end{align}
Here, $N$ is the number of $\bm{q}$ points in the first Brillouin zone, $n_{i} = n(\Omega_{q_{i}})$, $\omega_{c} = \omega + i0^{+}$ with $0^{+}$ being a positive infinitesimal,
and the summation over $(q_{1}, q_{2})$ are restricted to the pairs satisfying the momentum conservation;
$\bm{q}\pm\bm{q}_{1}=\bm{q}_{2}+\bm{G}$. 
For the ``opt.'' case, the $\Delta_{q}$ value of the $T_{2g}$ mode calculated with the 9$\times$9$\times$9 $\bm{q}$ point mesh was
$-0.2$ cm$^{-1}$ at 0 K and $-1.1$ cm$^{-1}$ at 300K.
Hence, the experimental $T$-dependence of the Raman shift can be better explained with the inclusion of the effect of the intrinsic frequency shift by the cubic anharmonicity. 
We have also calculated the $\Delta_{q}$ values for the other systems with different cage sizes and found 
that the frequency shift by the bubble diagram is negative ($\Delta_{q} < 0)$
and less significant than the hardening by the quartic anharmonicity~\cite{supplement}.

\begin{figure}[t]
\centering
\includegraphics[width=8.5cm, clip]{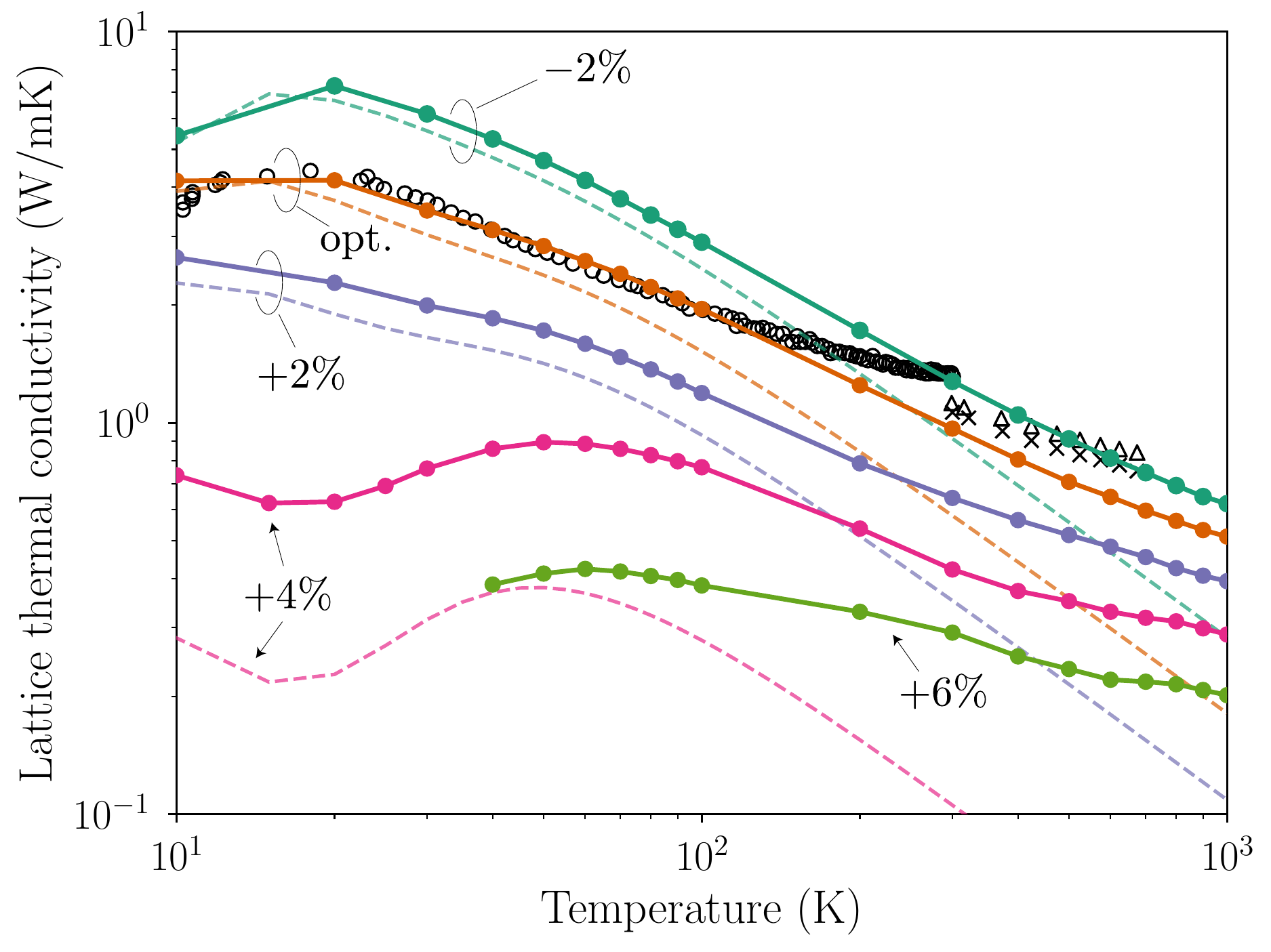}
\caption{(Color online) Lattice thermal conductivity of BGG calculated with various lattice constants.
The solid and dashed lines are obtained with the SCP+BTE method and the conventional BTE method, respectively. 
The experimental values in the low- ($\circ$) and high-temperature ($\times$, $\triangle$) regions are adapted from Refs.~\onlinecite{2001PhRvB..63x5113S,PhysRevB.80.125205}, respectively.}
\label{fig:kappa}
\end{figure}

To elucidate the intrinsic effects of the frequency renormalization on LTC quantitatively, 
we have calculated $\kappa_{L}$ based on the Boltzmann transport equation (BTE)
within the relaxation-time approximation, where
\begin{equation}
\kappa_{L}^{\mu\nu}(T) = \frac{1}{NV}\sum_{q} c_{q}(T)v^{\mu}_{q}(T)v^{\nu}_{q}(T)\tau_{q}(T).
\label{eq:kappa}
\end{equation}
Here, $V$ is the unit cell volume, $c_{q}(T)$ is the mode specific heat, 
$\bm{v}_{q}(T)$ is the group velocity, and $\tau_{q}(T)$ is the lifetime of phonon $q$.
Unlike the conventional BTE approach, where harmonic frequencies and eigenvectors are used as the ground state 
for calculating $\bm{v}_{q}$ and $\tau_{q}$, we employ the SCP frequencies and eigenvectors in the present
SCP+BTE method [Eq.~(\ref{eq:kappa})]. 
Therefore, the group velocity $\bm{v}_{q}(T)=\partial \Omega_{q}(T)/\partial \bm{q}$ also shows intrinsic $T$-dependence.
The phonon lifetime $\tau_{q}$ is estimated using the Matthiessen's rule 
$\tau_{q}^{-1} = \tau_{q,\mathrm{anh}}^{-1} + \tau_{q,\mathrm{iso}}^{-1} + \tau_{q,\mathrm{b}}^{-1}$.
The anharmonic scattering rate is calculated from the imaginary part of the bubble self-energy [Eq.~(\ref{eq:bubble})]
as $\tau_{q,\mathrm{anh}}^{-1} = 2\Gamma_{q}^{(\mathrm{B})} = 2\mathrm{Im}\Sigma_{q}^{(\mathrm{B})}(\Omega_{q})$,
and the phonon-isotope scattering rate $\tau_{q,\mathrm{iso}}^{-1}$ is evaluated perturbatively~\cite{Tamura_iso}.
For the phonon-boundary scattering rate, we employ $\tau_{q,\mathrm{b}}^{-1} = 2|\bm{v}_{q}|/L$ with the grain size of $L = 2.5$ $\mu$m 
that reproduces the experimental crystalline peak of LTC~\cite{2001PhRvB..63x5113S}.

\begin{figure}[t]
\centering
\includegraphics[width=8.5cm, clip]{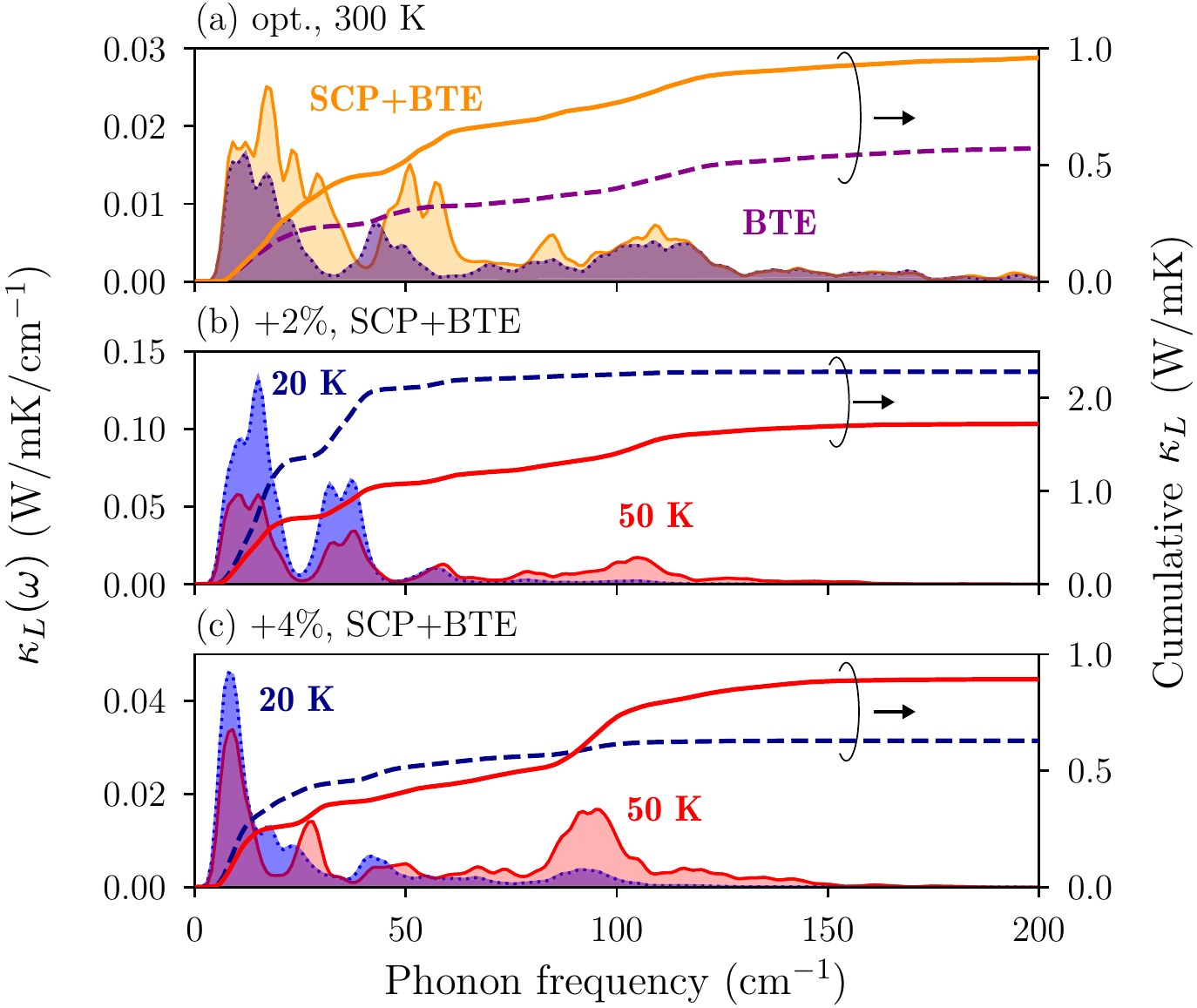}
\caption{(Color online) Thermal conductivity spectrum $\kappa_{L}(\omega)$ and its cumulative value.
(a) Comparison of the conventional BTE and SCP+BTE results for the optimized lattice constant at 300 K.
(b), (c) SCP+BTE results for the ``$+2$\%'' and ``$+4$\%'' systems compared at two different temperatures.}
\label{fig:kl_spec}
\end{figure}

Figure \ref{fig:kappa} shows $(V,T)$-dependence of LTC calculated by 
the BTE and the SCP+BTE methods with 9$\times$9$\times$9 $\bm{q}$ points. 
The LTC values calculated by the SCP+BTE method are generally higher than those obtained by the conventional BTE method.
This tendency becomes more pronounced with increasing $V$ and $T$.
The predicted $\kappa_{L}$ value by the SCP+BTE method is 0.97 W/mK at 300 K, which agrees well with the experimental values of 1.31 W/mK (Sales \textit{et al.}, Ref.~\onlinecite{2001PhRvB..63x5113S}) and 1.06 W/mK (May \textit{et al}., Ref.~\onlinecite{PhysRevB.80.125205}).
In contrast, the conventional method gives 0.58 W/mK, which is 40\% smaller than the SCP+BTE value.
Moreover, the deviation from $\kappa_{L} \propto T^{-1}$ in a high temperature range can be well reproduced by the SCP+BTE approach.
These results clearly reveal the essential role of the frequency renormalization on the thermal transport properties of BGG.
To elucidate the origin of the increase in $\kappa_{L}$ due to the hardening of the guest modes, 
we compare the LTC spectrum $\kappa_{L}(\omega)$ for the ``opt.'' case at 300 K.
The results are shown in Fig.~\ref{fig:kl_spec}(a).
With the hardening of the low-lying guest modes, $\kappa_{L}(\omega)$ increases significantly in the frequency region below 70 cm$^{-1}$.
After careful investigation, we found that this increase can be ascribed to the decrease in the phonon linewidth $\Gamma_{q}^{(\mathrm{B})}$, whose magnitude is roughly proportional to the available scattering phase space (SPS) and the strength of the cubic coupling $|V^{(3)}(-q;q_{1};q_{2})|^{2}= |\Phi(-q;q_{1};q_{2})|^{2}/8\Omega_{q}\Omega_{1}\Omega_{2}$. In the frequency range of 45--65 cm$^{-1}$, the SPS decreases by $\sim 50$\% due to the hardening of the low-frequency guest modes~\cite{supplement}.
This reduction results in the enhancement of $\kappa_{L}(\omega)$ in the same frequency range.
In the low-frequency region below 45 cm$^{-1}$, however, the change of the SPS is too small to explain 
the corresponding enhancement of $\kappa_{L}(\omega)$. 
Indeed, we found that the coupling coefficient $|V^{(3)}(-q;q_{1};q_{2})|^{2}$ is also suppressed strongly due to the
reduction of the optical-acoustic phonon hybridization caused by the hardenings of the optical modes (see the SI~\cite{supplement}).
The reduction of $\Gamma_{q}^{(\mathrm{B})}$ by the anharmonic renormalization observed here highlights the importance of 
the (effective) harmonic force constants, in accord with the previous numerical and experimental studies~\cite{Pailhes:2014fe,Li:2016bg,PhysRevB.94.054310}.
We have also found that, with the inclusion of the effect of the frequency renormalization, the theoretical $\Gamma_{q}^{(\mathrm{B})}$ values of TA modes 
agree quantitatively with the experimental results of Lory \textit{et al.}~\cite{Lory:2017eh} at 300 K (see the SI~\cite{supplement}).

Next, we discuss the volume dependence of the LTC calculated by SCP+BTE.
As can be inferred from Fig.~\ref{fig:kappa}, the LTC values decrease with increasing the unit cell volume.
It is interesting to observe that the crystalline peak of $\kappa_{L}$ near 20 K evolved into a resonance dip when the  
the lattice constant was expanded up to $+4$\%. 
In the ``$+4$\%'' system, $\kappa_{L}$ takes the minimum value at $\sim$15 K and increases up to 50 K, which is qualitatively different from the $T$-dependence of the other systems.
These computational results agree qualitatively with the experimental results on Sr$_{8}$Ga$_{16}$Si$_{30-x}$Ge$_{x}$, where
the cage size was controlled by changing the $x$ value~\cite{2007PhRvB..75s5210S}.
To understand the microscopic origin of the increase of $\kappa_{L}$ from $\sim$15 K to 50 K,
we compare the calculated LTC spectra of the ``$+2$\%'' and ``$+4$\%'' systems in Figs.~\ref{fig:kl_spec}(b) and (c), respectively.
In the both systems, the dispersive phonon modes below 50 cm$^{-1}$ contribute more than 90\% to the total $\kappa_{L}$ value at 20 K. 
When higher frequency phonon modes are thermally excited at 50 K via the enhancement of $c_{q}(T)$, 
the contribution from the dispersive optical modes around 100 cm$^{-1}$ becomes significant.
In the ``$+2$\%'' system, the thermal enhancement of $\kappa_{L}(\omega)$ near 100 cm$^{-1}$ was smaller than
the concurrent reduction in $\omega < 50$ cm$^{-1}$, which resulted in 
$\kappa_{L}(20 \mathrm{K}) > \kappa_{L}(50 \mathrm{K})$.
When the lattice constant is changed from ``$+2$\%'' to ``$+4$\%'', the frequencies of the low-lying guest modes
decrease as discussed earlier. This softening causes suppression of $\kappa_{L}(\omega)$, which is particularly significant in $\omega < 50$ cm$^{-1}$ while the phonon lifetimes as well as $\kappa_{L}(\omega)$ in the higher frequency region ($\sim$100 cm$^{-1}$) are less affected~\cite{supplement}. 
Such a joint effect of strong acoustic-optical scatterings in the low-frequency range and the presence of higher-frequency dispersive 
optical modes
resulted in $\kappa_{L}(20 \mathrm{K}) < \kappa_{L}(50 \mathrm{K})$ for the ``$+4$\%'' system.
It is important to mention that the predicted $\kappa_{L}$ values in the low-$T$ region depends on the 
employed grain size $L$. 
We found that the resonance dip of the ``$+4$\%'' system persisted even when we omitted the boundary scattering term.
Moreover, the resonance dip changed into a plateau when we employed a smaller $L$ value,
which is consistent with the experimental fact that the presence of the resonance dip is sensitive to details of sample
preparation methods~\cite{1998ApPhL..73..178N,2001PhRvB..63x5113S,Christensen:2016dv}.

Finally, we focus on the $T$-dependence of $\kappa_{L}$ above 50 K in Fig.~\ref{fig:kappa}.
The $T$-dependence obtained by the conventional BTE method follows $\kappa_{L} \propto T^{-1}$ for 
all of the studied systems, which do not agree with the experimental fact that $\kappa_{L}$ shows weaker 
$T$-dependence~\cite{PhysRevB.80.125205,2007PhRvB..75s5210S}.
The SCP+BTE method considerably improved the agreement with the experimental results and produced milder temperature dependence.
However, it was still inadequate to reproduce the increasing trend of $\kappa_{L}$ observed in some clathrates having large guest-free-spaces.
Since the SCP+BTE method is still based on the phonon-gas model, it does not include the
nondiagonal Peierls contribution~\cite{1984PhRvB..29.2884A,Allen:1993gt} and the anharmonic contribution~\cite{2010PhRvB..82v4305S} to the 
heat flux operator, which become generally more important at high temperature.   
Therefore, our results also indicate that these contributions may not be negligible in some clathrates at high temperature.

To summarize, we performed fully \textit{ab initio} calculations of $\kappa_{L}$ of type-I clathrate BGG with various different cage sizes,
where both the three-phonon scattering and the frequency renormalization by the quartic anharmonicity were taken into account.
We showed that the hardening of vibrational frequencies of rattling modes caused by the quartic anharmonicity
significantly affects the calculated $\kappa_{L}$ values, leading to an improved agreement with experimental results
including the $T$-dependence weaker than $\kappa_{L} \propto T^{-1}$.
In addition, we found that the evolution from crystal-like to \textit{glasslike} $\kappa_{L}$ near $\sim$20 K
can be realized by our static calculations without disorders, which can be
attributed to the presence of low-frequency guest modes that strongly couple with acoustic modes along with
higher-frequency dispersive optical modes and the phonon-boundary scatterings.
While our simulation does not exclude the possibility of disorders to further reinforce the \textit{glasslike} behavior 
observed in real clathrate samples, it provides a new miscroscopic insight into the exceptional thermal transport of clathrates.

\acknowledgements

We thank Koichiro Suekuni and St{\'e}phane Pailh{\`e}s for fruitful discussions, and Kiyoyuki Terakura for useful comments.
This study is partly supported by JSPS KAKENHI Grant Number 16K17724, ``Materials research by Information Integration'' Initiative (MI2I) project
of the Support Program for Starting Up Innovation Hub from Japan Science and Technology Agency (JST),
and MEXT Element Strategy Initiative to Form Core Research Center in Japan.
The computation in this work has been done using the facilities of the Supercomputer Center,
Institute for Solid State Physics, The University of Tokyo.

\clearpage
\setcounter{equation}{0}
\setcounter{figure}{0}
\setcounter{table}{0}
\setcounter{page}{1}
\makeatletter
\renewcommand{\theequation}{S\arabic{equation}}
\renewcommand{\thefigure}{S\arabic{figure}}
\renewcommand{\thetable}{S\arabic{table}}
\renewcommand{\thesection}{S\arabic{section}}
\renewcommand{\bibnumfmt}[1]{[S#1]}
\renewcommand{\citenumfont}[1]{S#1}

\begin{center}
\textbf{\large Supplemental Materials \\ {\small ``Quartic Anharmonicity of Rattlers and Its Effect on Lattice Thermal Conductivity of Clathrates from First Principles''}}
\end{center}

\section{A. Details of computational procedure}

In this study, we performed DFT calculations using the Vienna \textit{Ab initio} simulation package (\textsc{vasp})~\cite{Kresse1996} with the
projector augmented wave (PAW) method~\cite{PAW1994,Kresse1999}. 
We employed the cutoff energy of 320 eV and 4$\times$4$\times$4 Monkhorst-Pack $k$ grid~\cite{MonkhorstPack} for the Brillouin zone integration. 
For the exchange-correlation functional, we employed the Perdew-Burke-Ernzerhof (PBE) functional~\cite{PBE1996}.
As a structural model for Ba$_{8}$Ga$_{16}$Ge$_{30}$, we employed a fully ordered unit cell as in Ref.~\cite{PhysRevLett.114.095501}, where the Ga atoms occupy the $16i$ wyckoff sites.
Before performing phonon calculations, we fully relaxed the lattice constant $a$ and the internal coordinates, where we obtained $a = 10.954$~\AA.
Then, we changed the lattice constant from the optimized value by $-2$\%, $+2$\%, $+4$\%, and $+6$\%, and relaxed the internal coordinates for them. 
The actual values of the lattice constants and the resulting pressures are shown in table~\ref{table:alat}.

\begin{table}[b]
\caption{Lattice constants of Ba$_{8}$Ga$_{16}$Ge$_{30}$ employed in study along with the resulting external pressure.}
\label{table:alat}
\begin{ruledtabular}
\begin{tabular}{ccc}
Label & Lattice constant (\AA) & Pressure (GPa) \\ \hline
$-2$\% & 10.735 & 3.5 \\
opt. & 10.954 & 0.0 \\
$+2$\% & 11.173 & $-2.6$ \\
$+4$\% & 11.392 & $-4.5$ \\
$+6$\% & 11.611 & $-5.8$ 
\end{tabular}
\end{ruledtabular}
\end{table}

After the relaxation, we calculated harmonic interatomic force constants (IFCs) using the finite displacement approach~\cite{1997PhRvL..78.4063P}.
Here, we considered all harmonic IFCs inside the unit cell and employed the displacement length of $\Delta u = 0.02$ \AA.
The anharmonic IFCs were calculated using the compressive sensing lattice dynamics method~\cite{PhysRevLett.113.185501}.
To this end, we employed the following procedure: 
First, we performed \textit{ab initio} molecular dynamics (AIMD) for the ``$+4$\%'' system at 300 K and generated physically relevant atomic
configurations. 
The AIMD simulation was performed with 2$\times$2$\times$2 Monkhorst-Pack $k$ grid and a relatively large convergence criterion for the SCF loop to
accelerate the structural sampling. 
Second, we uniformly sampled 140 snapshots from the AIMD trajectory. 
In each sampled snapshot, we further displaced all atoms by 0.1 \AA{} in random directions to
decrease cross-correlations between the sampled configurations.
Third, we performed static DFT calculations for the 140 snapshots and calculated Hellmann-Feynman forces accurately.
These displacement and force data sets form a training data for estimating anharmonic IFCs of the ``$+4$\%'' system.
For the other systems with different unit cell volumes, the atomic snapshots were created simply by changing the lattice constants while
leaving the internal coordinates of the original structures unaltered.
In total, $140 \times 5 = 700$ DFT calculations were conducted for calculating anharmonic IFCs of Ba$_{8}$Ga$_{16}$Ge$_{30}$.
Finally, we estimated anharmonic IFCs by using the least absolute shrinkage and selection operator (LASSO)~\cite{Tibshirani:1996wb}.
The hyperparameter for the LASSO was selected from cross-validation scores~\cite{Hastie:2009fg}, 
and the optimization problem was solved by using the coordinate descent method~\cite{Hastie:2015tb}.

In the present study, we considered anharmonic IFCs up to the sixth order. 
To reduce the number of independent IFCs, we introduced cutoff radii for anharmonic terms so that only
the host-guest (host-host) terms inside the first (second) nearest neighbor shell were included.
In addition, we omitted less important four-body terms for the quartic IFCs and multi-body ($> 2$ body) terms for the quintic and sextic IFCs.
After removing linearly-dependent IFCs by using available symmetry operations and the constraints of the translational invariance,
the numbers of irreducible IFCs turned out to be 994, 2819, 252, and 345 for cubic, quartic, quintic, and sextic terms, respectively.  
We estimated these 4410 terms by solving the $\ell_{1}$-regularized linear regression problems.
The calculated cubic and quartic IFCs were employed for the subsequent phonon calculations.

The self-consistent phonon (SCP) calculations were performed at $\bm{q} = \bm{q}_{1} = \bm{0}$, and anharmonic phonon frequencies and eigenvectors at
arbitrary $\bm{q}$ points were obtained by Fourier interpolation.　
For thermal conductivity calculations, we employed the gamma-centered 9$\times$9$\times$9 $\bm{q}$ points, with which we obtained well-converged results.
The Dirac delta function appearing in $\Gamma_{q}^{(\mathrm{B})}$ was evaluated using the tetrahedron method~\cite{PhysRevB.49.16223}.
All of the phonon calculations and IFC estimations were performed by using the \textsc{alamode} package~\cite{Tadano2014,alamode}.

\section{B. Anharmonic phonon dispersion}

\begin{figure*}
\centering
\includegraphics[width=0.8\textwidth,clip]{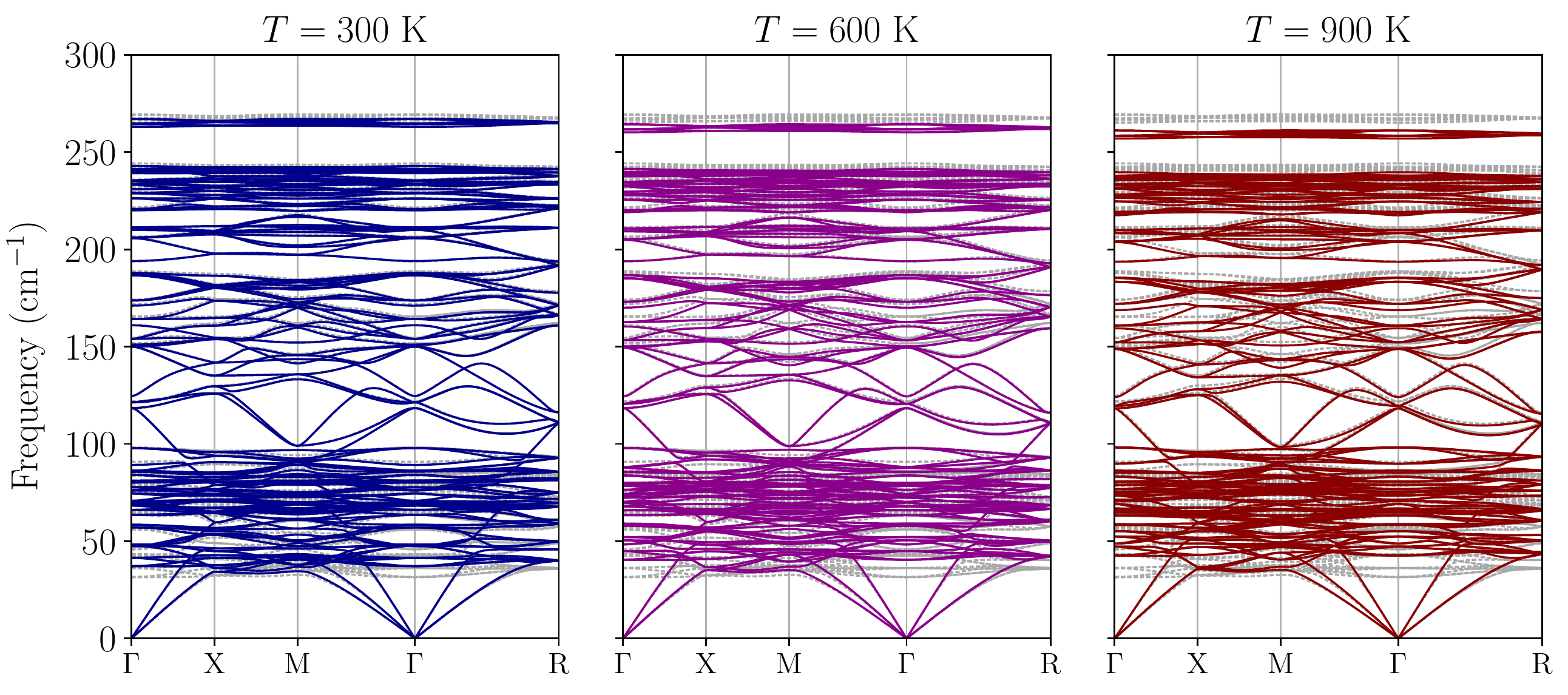}
\caption{Anharmonic phonon dispersion curves of Ba$_{8}$Ga$_{16}$Ge$_{30}$ along high-symmetry lines of the Brillouin zone.
The results for the ``$-2$\%'' system are shown.
The SCP dispersion curves at 300 K, 600 K, and 900 K are shown by solid lines in the left, middle, and right panels, respectively.
The harmonic dispersion curves are also shown by gray dotted lines.}
\label{fig:SCP1}
\end{figure*}

\begin{figure*}
\centering
\includegraphics[width=0.8\textwidth,clip]{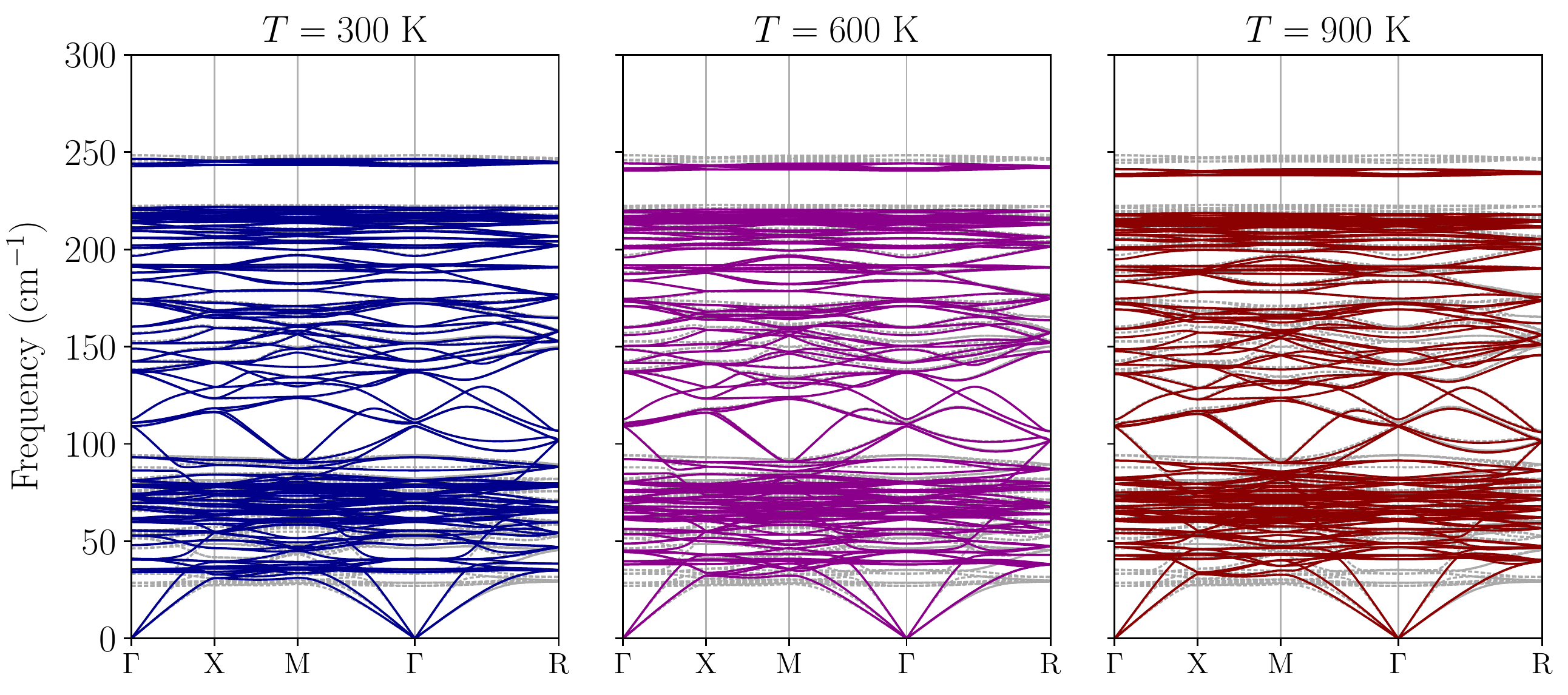}
\caption{The results for the ``opt.'' system. See Fig.~\ref{fig:SCP1} caption for explanation.}
\label{fig:SCP2}
\end{figure*}

\begin{figure*}
\centering
\includegraphics[width=0.8\textwidth,clip]{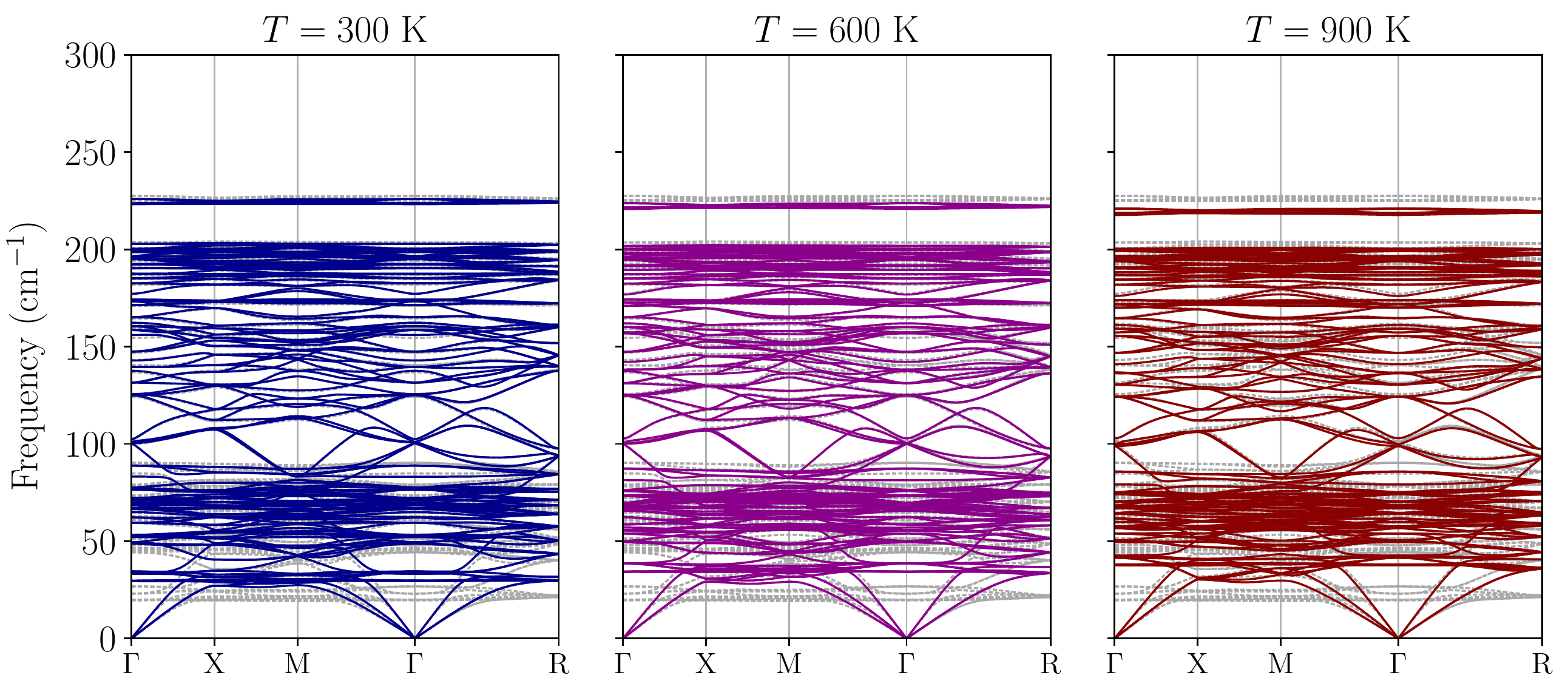}
\caption{The results for the ``$+2$\%'' system. See Fig.~\ref{fig:SCP1} caption for explanation.}
\label{fig:SCP3}
\end{figure*}

\begin{figure*}
\centering
\includegraphics[width=0.8\textwidth,clip]{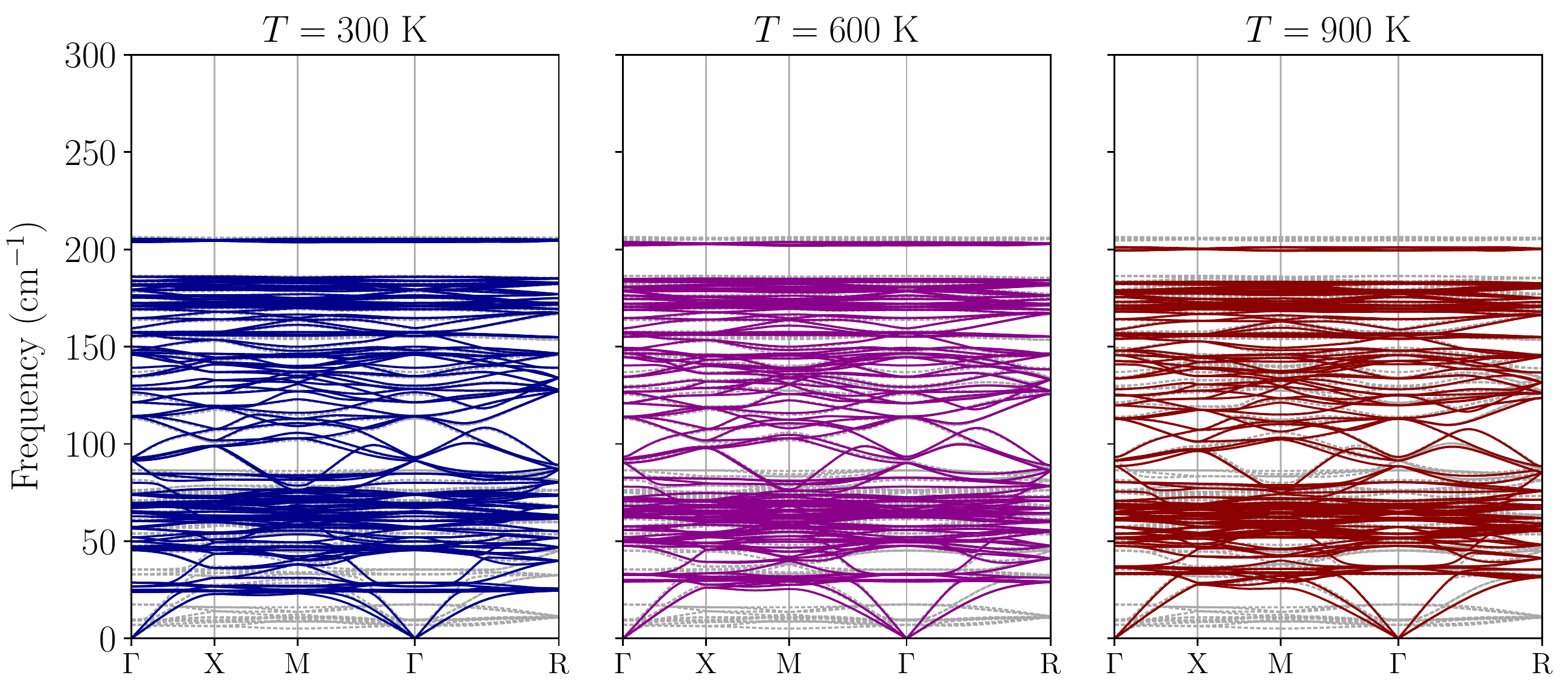}
\caption{The results for the ``$+4$\%'' system. See Fig.~\ref{fig:SCP1} caption for explanation.}
\label{fig:SCP4}
\end{figure*}

\begin{figure*}
\centering
\includegraphics[width=0.8\textwidth,clip]{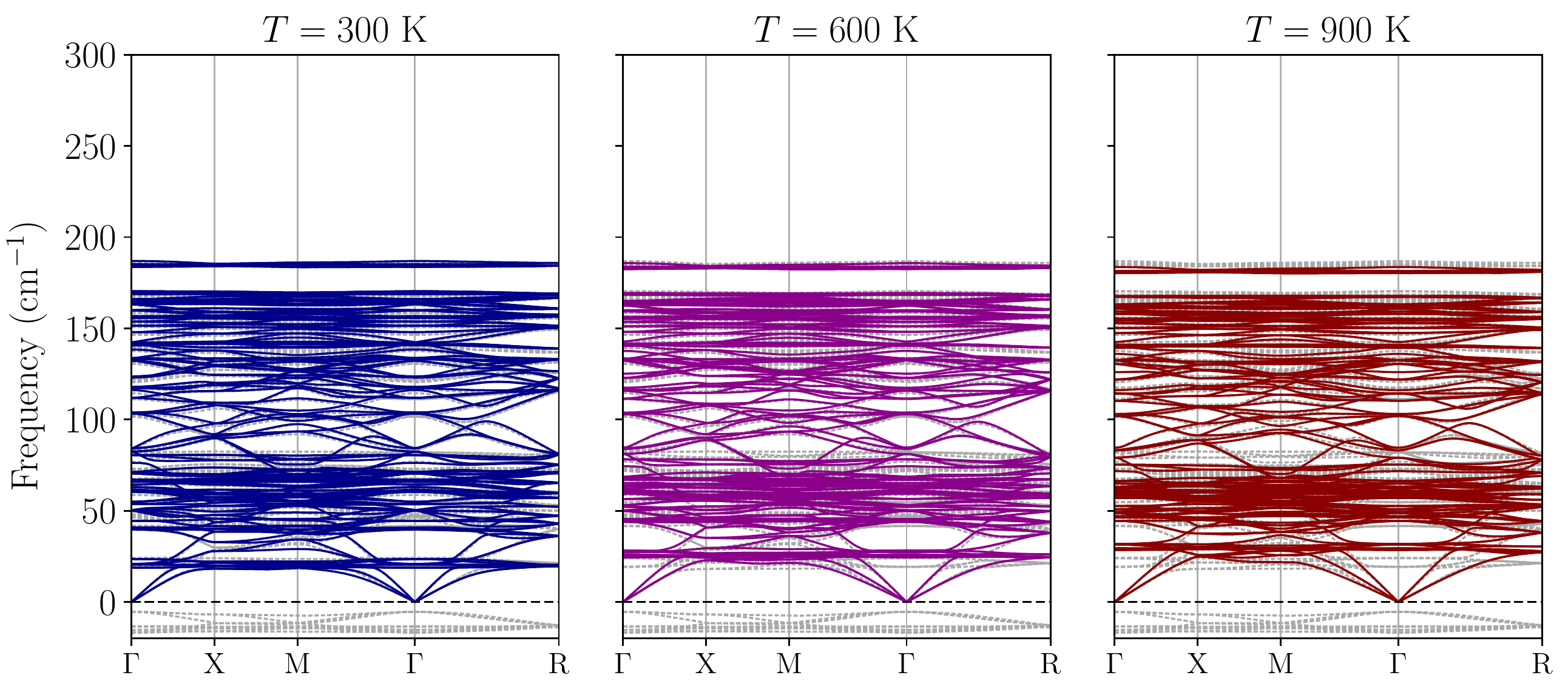}
\caption{The results for the ``$+6$\%'' system. See Fig.~\ref{fig:SCP1} caption for explanation.}
\label{fig:SCP5}
\end{figure*}

The full anharmonic phonon dispersion curves calculated by the self-consistent phonon (SCP) theory are shown in Figs.~\ref{fig:SCP1}--\ref{fig:SCP5}.
In these figures, we also show harmonic phonon dispersion curves for comparison.

\section{C. Frequency shift by cubic anharmonicity}

In table \ref{table:frequency}, the frequencies of the $T_{2g}$ guest mode calculated with different conditions and approximations are summarized.
The ``SCP+Bubble'' corresponds to the value of $\Omega_{q}+\Delta_{q}$ where $\Delta_{q} = -\mathrm{Re}\Sigma_{q}^{(\mathrm{B})}(\Omega_{q})$.
For the ``$+6$\%'' system, we could not reach a convergence of the SCP equation at 0 K within a reasonable computational time.
This is reasonable because the present SCP method does not allow the spontaneous symmetry breaking to occur 
and the thermally averaged equilibrium positions of the rattlers are assumed to be the center of the cage irrespective of temperature.
For off-center systems, such an assumption is invalid in a low-temperature region, 
but should be more or less reasonable at high temperatures.

\begin{table}[h]
\caption{\label{table:frequency} Calculated harmonic and anharmonic phonon frequencies (cm$^{-1}$) 
of the $T_{2g}$ guest modes with various lattice constants and temperatures.}
\begin{ruledtabular}
\begin{tabular}{crrrrrr}
 & \multirow{2}{*}{$T$ (K)} & \multicolumn{5}{c}{Lattice constant} \\
 \cline{3-7}
 & & $-2$\% & opt. & $+2$\% & $+4$\% & $+6$\%  \\ 
\hline
 Harmonic   &     & 36.2 & 28.6 & 19.7 &  7.0 & 15.3$i$\\ \addlinespace[1.3ex]

 SCP        &   0 & 37.1 & 29.8 & 21.4 & 11.3 & N/A \\
            & 300 & 41.5 & 35.5 & 29.5 & 23.9 & 18.8 \\
            & 600 & 44.9 & 39.7 & 34.4 & 29.3 & 24.5 \\
            & 900 & 47.2 & 42.8 & 37.9 & 33.2 & 28.5 \\ \addlinespace[1.3ex]

 SCP+Bubble &   0 & 36.9 & 29.6 & 21.1 & 10.7 & N/A \\
            & 300 & 40.4 & 34.4 & 27.9 & 21.5 & 14.8 \\
            & 600 & 43.1 & 37.9 & 32.3 & 26.6 & 21.1 \\
            & 900 & 44.6 & 40.4 & 35.6 & 30.4 & 25.2 \\
\end{tabular}
\end{ruledtabular}
\end{table}


\section{D. Phonon lifetime}

\begin{figure*}
\centering
\includegraphics[width=0.8\textwidth,clip]{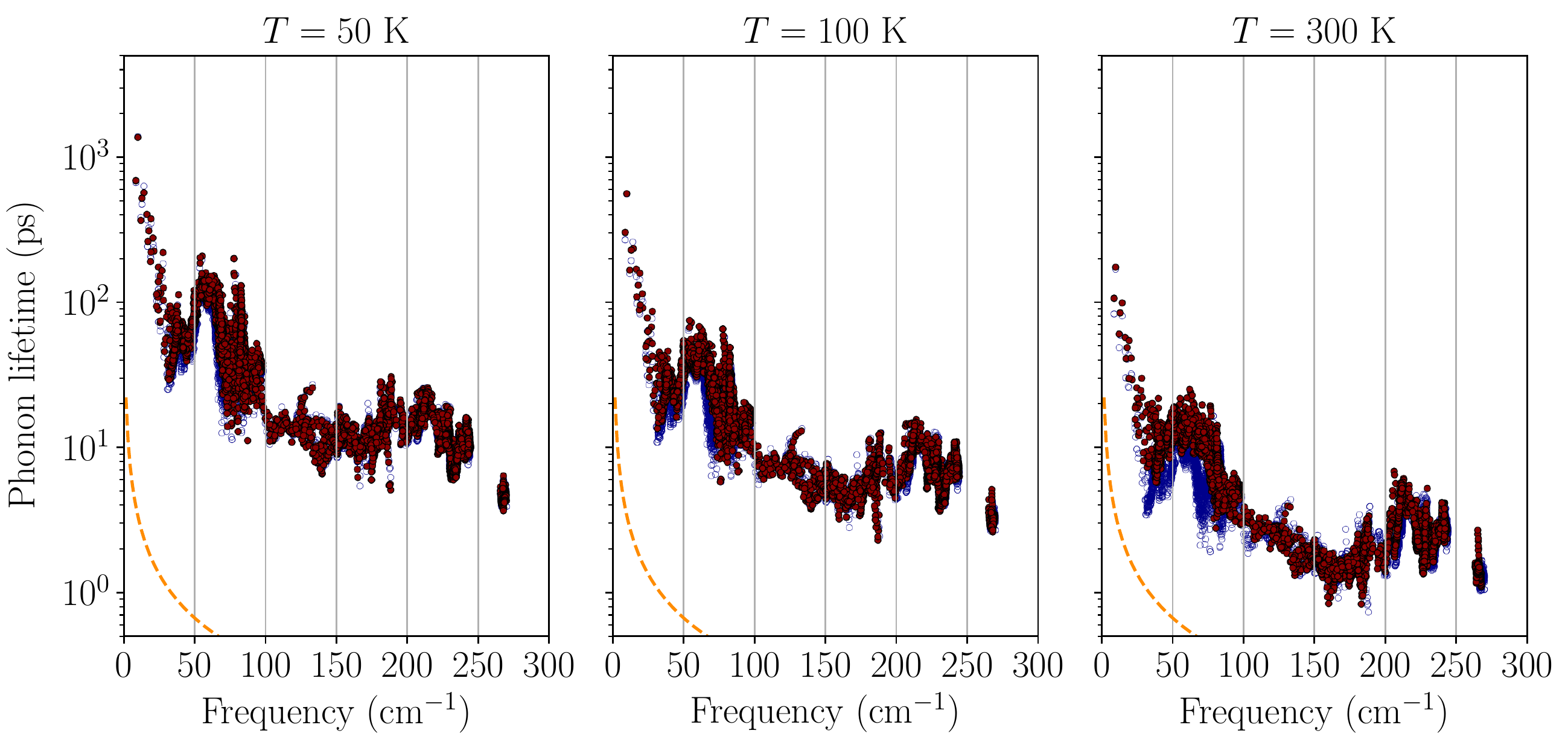}
\caption{Calculated phonon lifetimes of Ba$_{8}$Ga$_{16}$Ge$_{30}$.
The results for the ``$-2$\%'' system are shown.
The blue open (red filled) circles are results obtained by using the harmonic (SCP) lattice dynamics wavefunctions.
The results at 50 K, 100 K, and 300 K are shown in the left, middle, and right panels, respectively.
The orange dashed lines indicate $\tau = 2\pi\omega^{-1}$. 
The phonon quasiparticle picture becomes valid when the data points are well above this line.}
\label{fig:lifetime1}
\end{figure*}

\begin{figure*}
\centering
\includegraphics[width=0.8\textwidth,clip]{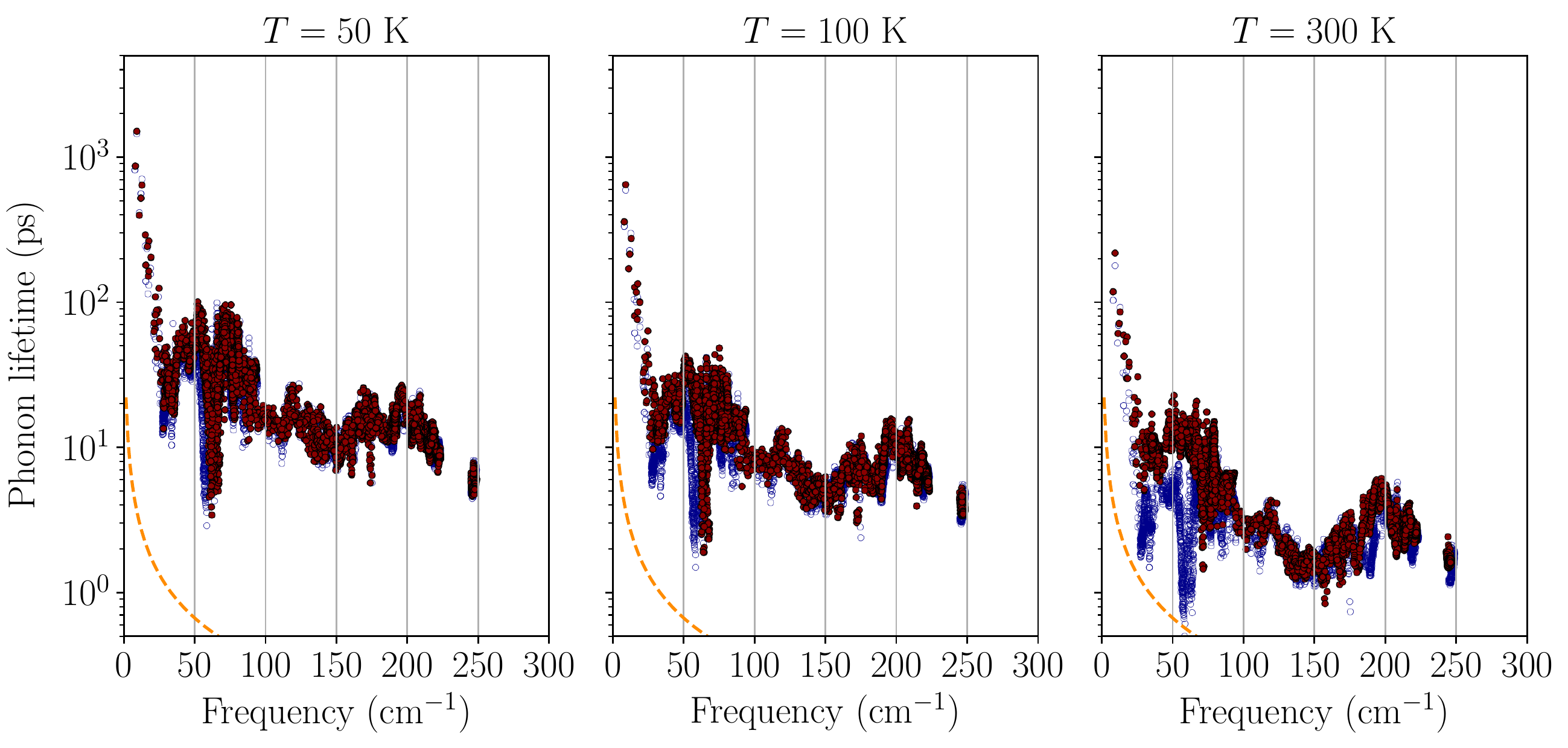}
\caption{The results for the ``opt.'' system. See Fig.~\ref{fig:lifetime1} caption for explanation.}
\label{fig:lifetime2}
\end{figure*}

\begin{figure*}
\centering
\includegraphics[width=0.8\textwidth,clip]{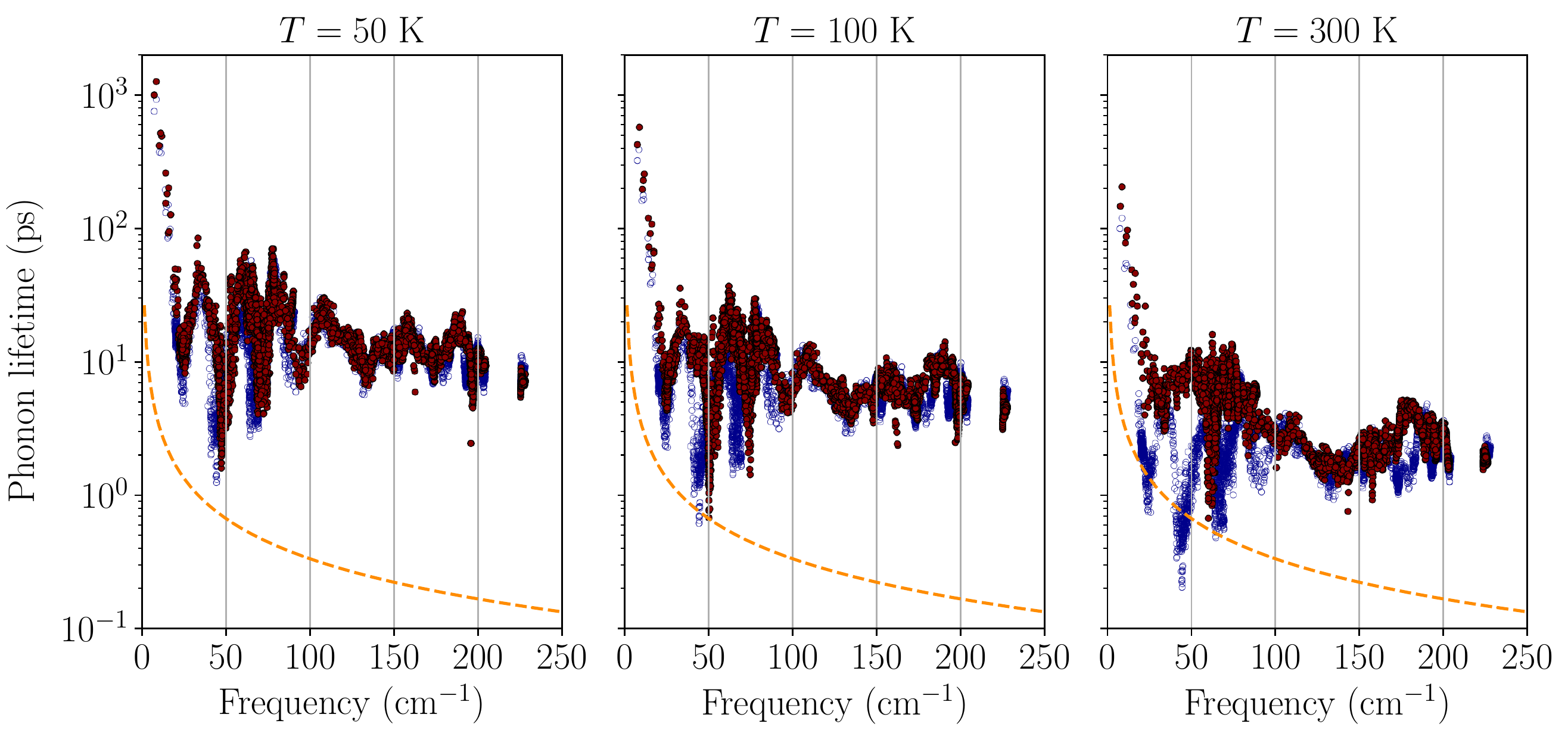}
\caption{The results for the ``$+2$\%'' system. See Fig.~\ref{fig:lifetime1} caption for explanation.}
\label{fig:lifetime3}
\end{figure*}

\begin{figure*}
\centering
\includegraphics[width=0.8\textwidth,clip]{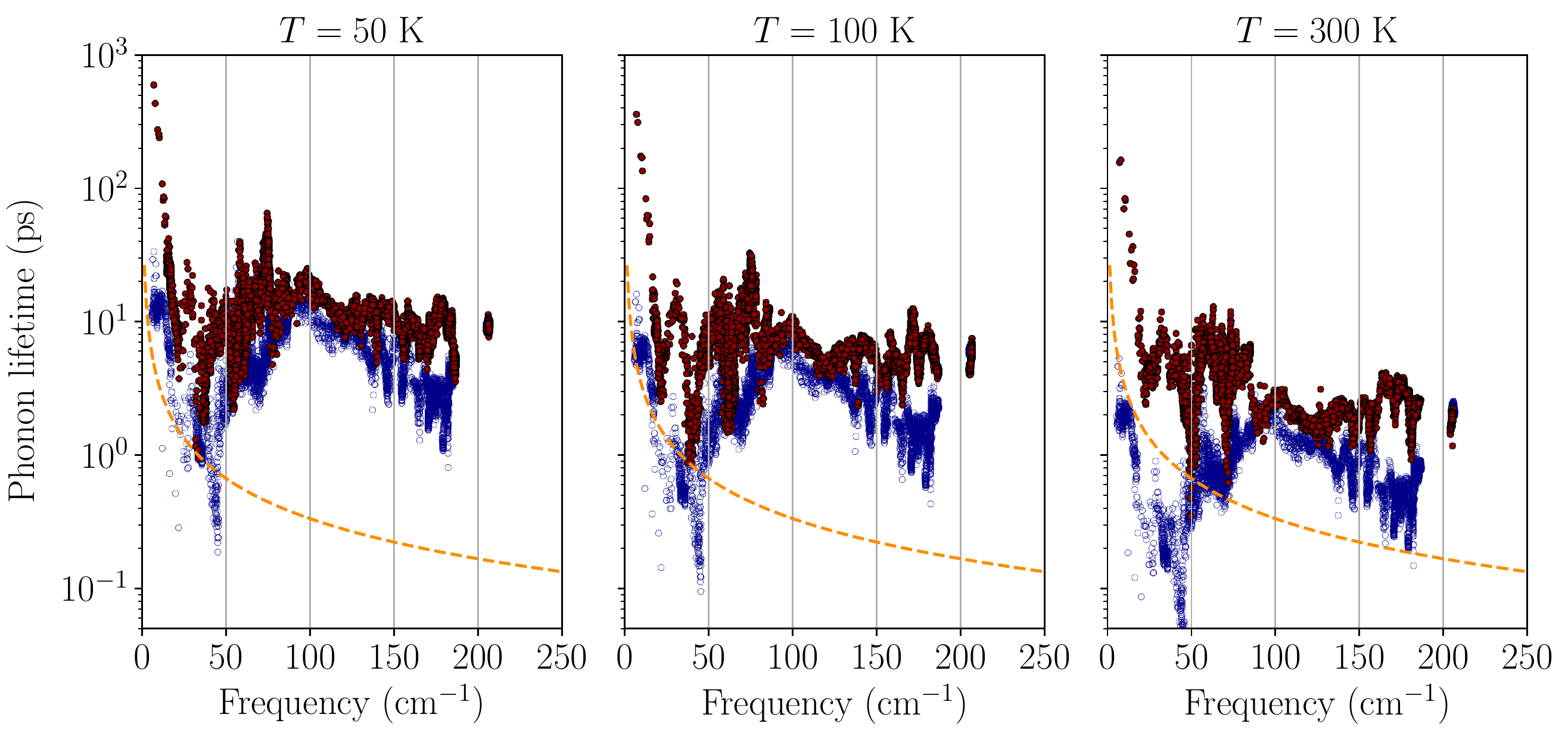}
\caption{The results for the ``$+4$\%'' system. See Fig.~\ref{fig:lifetime1} caption for explanation.}
\label{fig:lifetime4}
\end{figure*}

\begin{figure*}
\centering
\includegraphics[width=0.8\textwidth,clip]{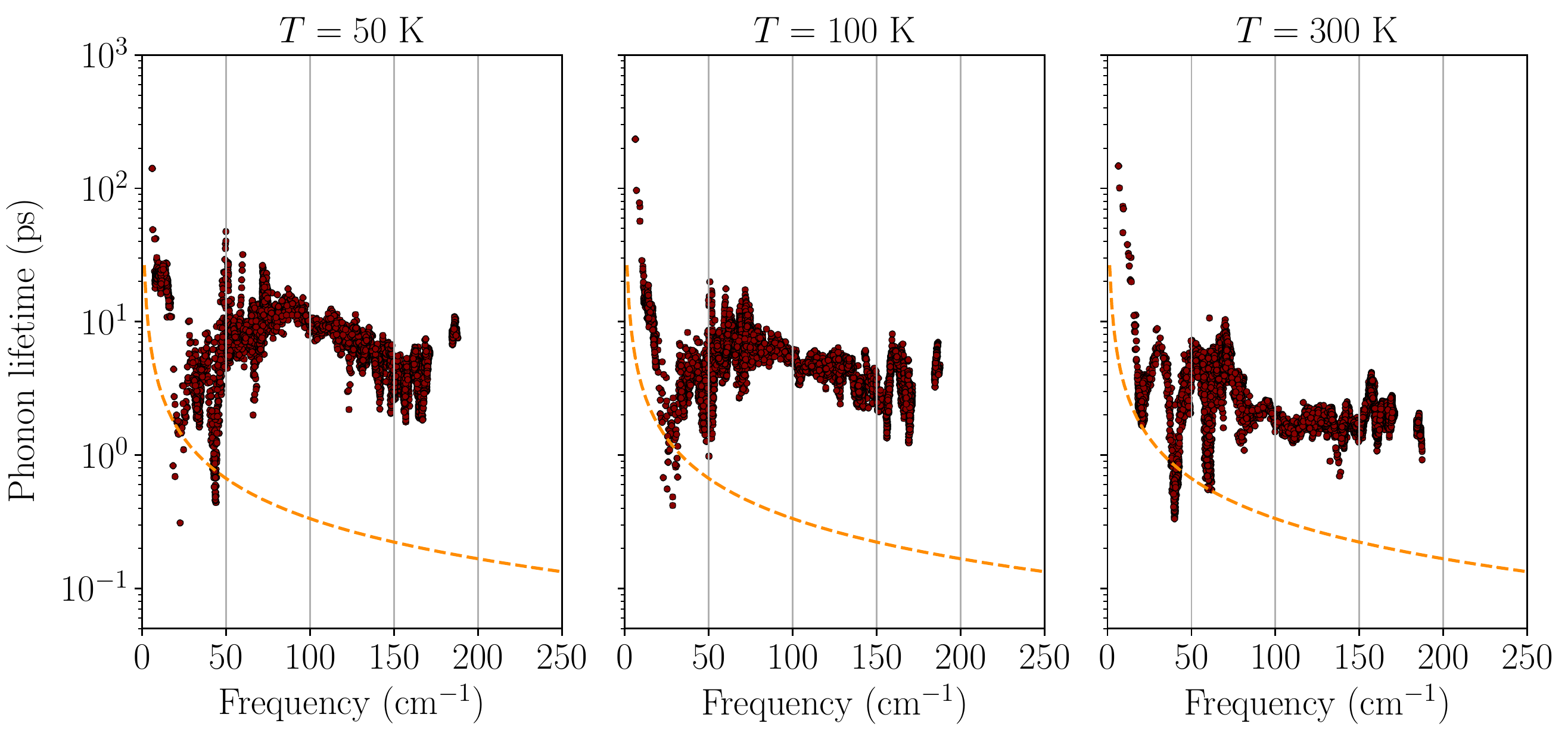}
\caption{The results for the ``$+6$\%'' system. See Fig.~\ref{fig:lifetime1} caption for explanation.}
\label{fig:lifetime5}
\end{figure*}

The calculated phonon lifetimes $\tau_{q,\mathrm{anh}} = \left(2\Gamma_{q}^{(\mathrm{B})}\right)^{-1}$ of Ba$_{8}$Ga$_{16}$Ge$_{30}$ are shown in Figs.~\ref{fig:lifetime1}--\ref{fig:lifetime5} as a function of phonon frequencies. 
In each figure, we compare two computational results; one is based on the harmonic approximation and the other is based on the SCP theory. For the ``$+6$\%'' system, we only show the result based on the SCP theory.
To see the validity of the quasiparticle picture, we also show $\tau= 2\pi\omega^{-1}$ by dashed lines.
When $\tau_{q} > 2\pi\omega_{q}^{-1}$ is not satisfied, the quasiparticle picture becomes questionable. 

\begin{figure*}[p]
\centering
\includegraphics[width=0.95\textwidth,clip]{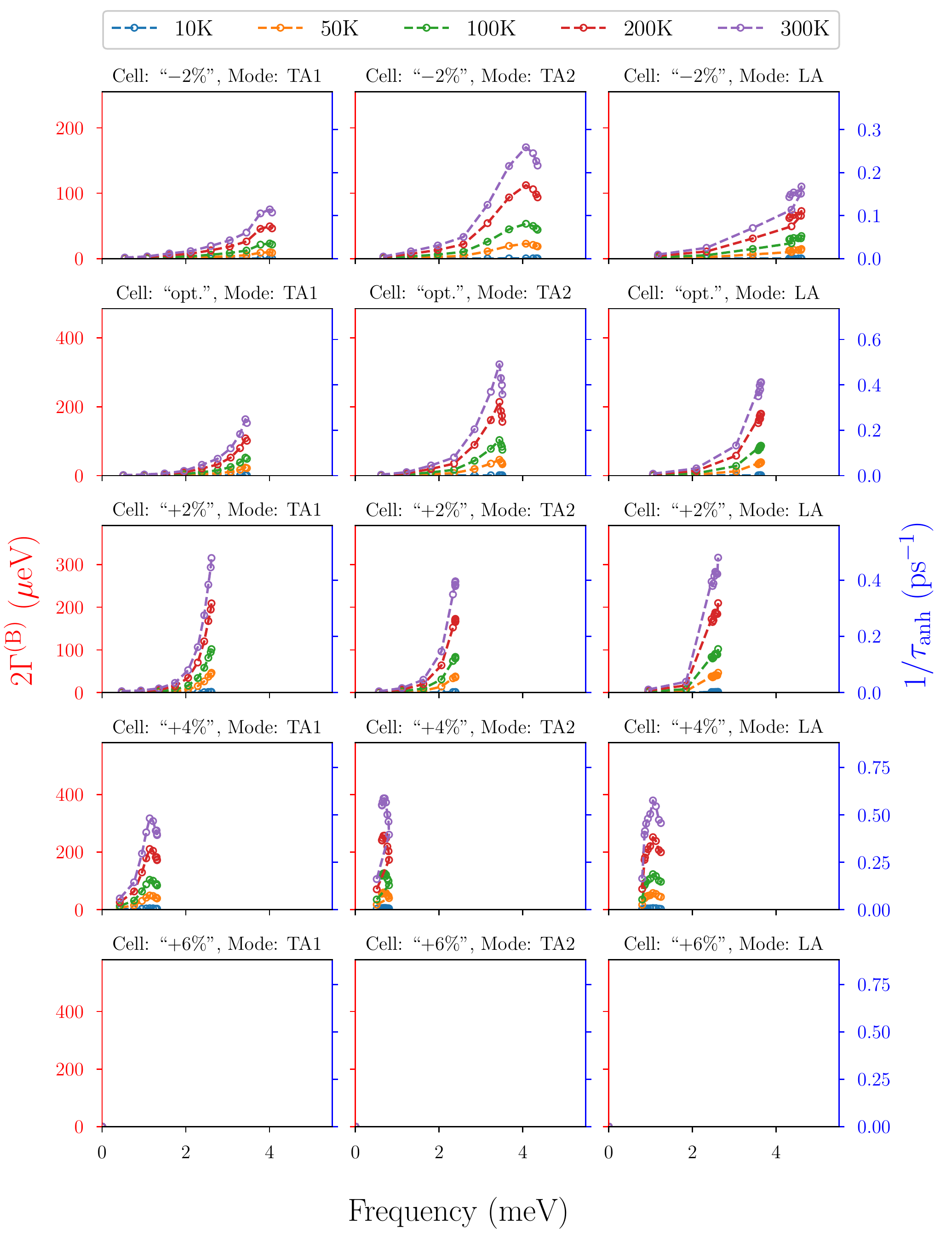}
\caption{The full width at half maximum (FWHM) of acoustics modes along the $[011]$ direction calculated by using harmonic phonon frequencies and eigenvectors.
The result for the ``$+6\%$'' system is not shown due to the existence of imaginary modes.}
\label{fig:Gamma_acoustic}
\end{figure*}

\begin{figure*}[p]
\centering
\includegraphics[width=0.95\textwidth,clip]{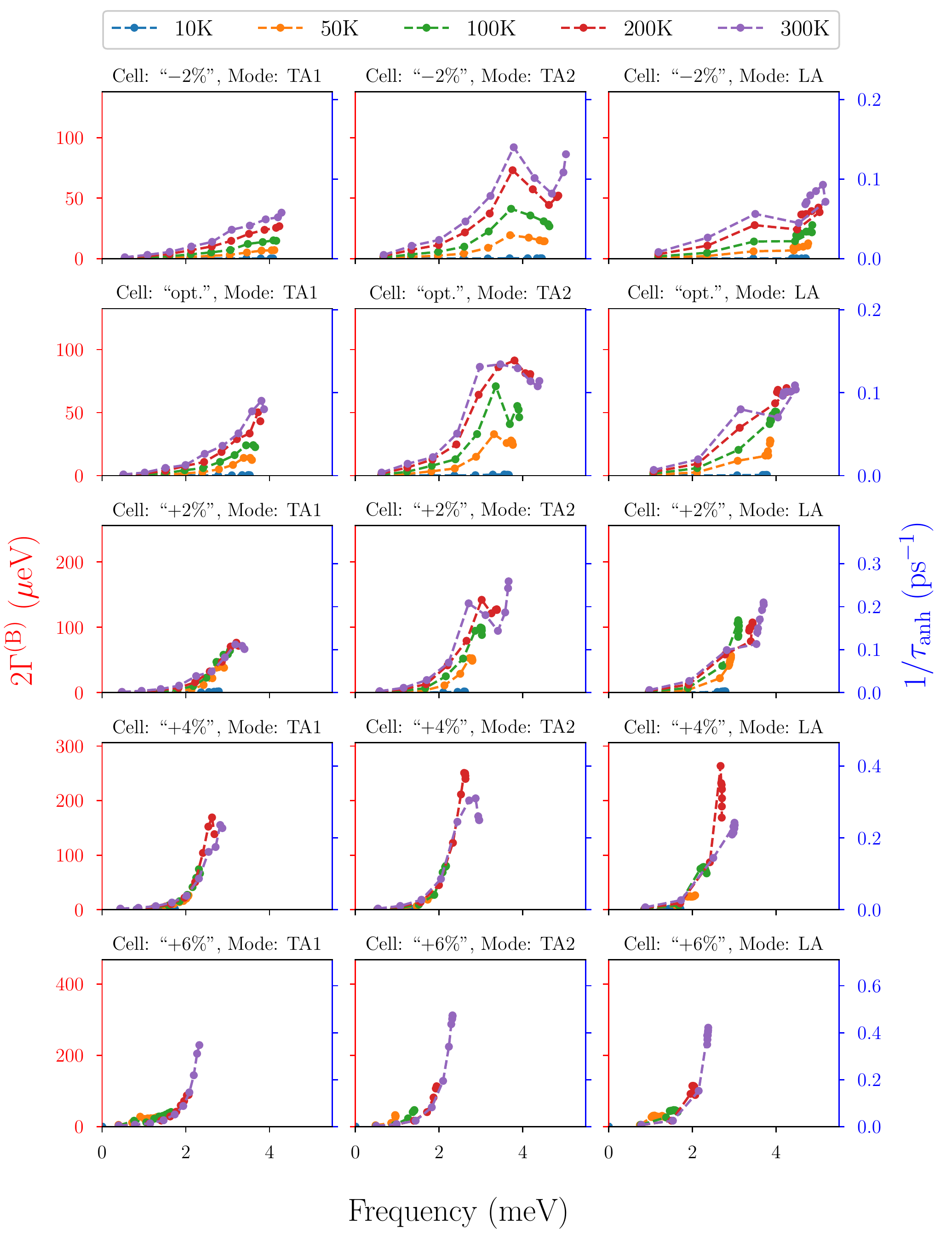}
\caption{The full width at half maximum (FWHM) of acoustics modes along the $[011]$ direction calculated by using the SCP frequencies and eigenvectors.}
\label{fig:Gamma_acoustic2}
\end{figure*}

Recently, Lory \textit{et al.} have determined the phonon lifetimes of TA modes of Ba$_{7.81}$Ge$_{40.67}$Au$_{5.33}$ ($a=10.85$ \AA) along the $[011]$ direction from the neutron resonant spin-echo (NRSE) spectrometer~\cite{Lory:2017eh}.
According to their study, the inverse phonon lifetime of the TA mode at 300 K was $1/\tau \approx $ 0.06--0.07 ps$^{-1}$ near the avoided-crossing region.
In Figs. \ref{fig:Gamma_acoustic} and \ref{fig:Gamma_acoustic2}, we show the calculated values of $2\Gamma^{(\mathrm{B})} = 1/\tau_{\mathrm{anh}}$ of acoustic modes along the $[011]$ direction. Here, we show the results of the acoustic modes at $\bm{q}_{i} = (0, \frac{i}{20}, \frac{i}{20})$ for $i = 1, 2, \dots, 10$. 
When the harmonic lattice dynamics wavefunction is used (Fig.~\ref{fig:Gamma_acoustic}), 
the $1/\tau_{\mathrm{anh}}$ value at 300 K for the ``opt.'' system, whose lattice constant is closest to that of Ba$_{7.81}$Ge$_{40.67}$Au$_{5.33}$, 
is as large as 0.25 and 0.49 ps$^{-1}$ for the TA1 and TA2 modes, respectively.
These results overestimate the experimental values by a factor 4 to 8, in accord with the DFT results of Lory~\cite{Lory:2017eh}.
When we considered the temperature dependence of phonon frequencies by the SCP theory (Fig.~\ref{fig:Gamma_acoustic2}), 
the $1/\tau_{\mathrm{anh}}$ values were reduced to 0.09 and 0.13 ps$^{-1}$ for the two TA modes.
Although the comparison here is not rigorous as we compare two different materials,
the results based on the SCP lattice dynamics wavefunctions agree well with experimental phonon lifetimes at 300 K,
thus indicating the importance of the anharmonic renormalization of phonon frequencies and eigenvectors.
When the temperature was decreased down to 10 K, the calculated $1/\tau_{\mathrm{anh}}$ values became as small as $\sim$0.001 ps$^{-1}$, whereas
the experimental value was $\sim$0.018 ps$^{-1}$.
The origin of the disagreement at 10 K may be attributed to the off-centering rattlers, 
additional scattering channels of phonons other than the three-phonon process, static and dynamical disorders, dislocations, 
and/or the resolution limit of the NRSE measurements.

\section{E. Change of scattering phase space and cubic coupling by frequency renormalization}

We show the change of the three-phonon scattering phase space (SPS) and the strength of the
cubic coupling coefficients caused by the hardening of low-lying guest modes. 
Here, all of the results were calculated with the optimized lattice constant and at 300 K.

\begin{figure}
\centering
\includegraphics[width=8.5cm,clip]{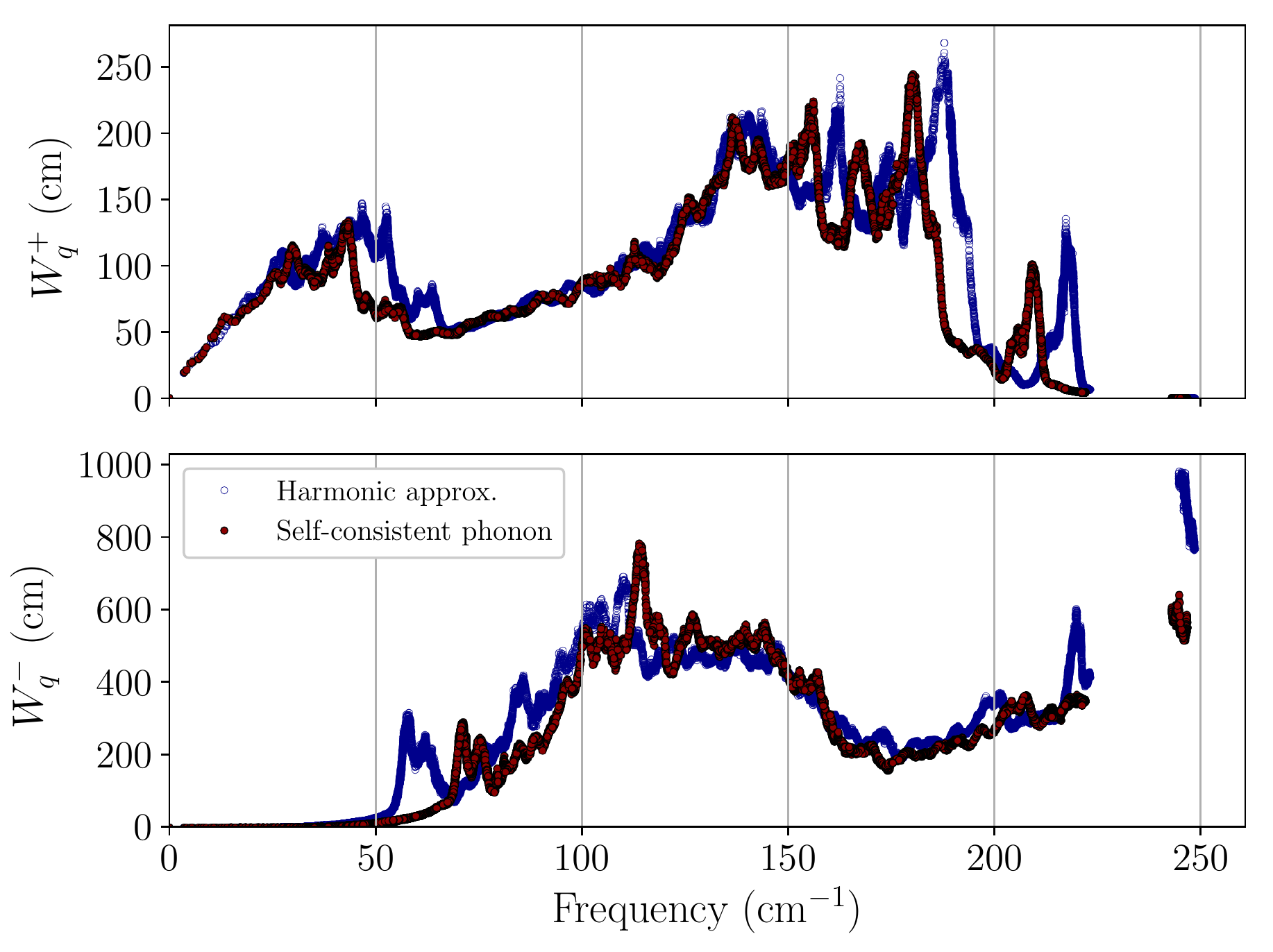}
\caption{The scattering phase space at 300 K shown as a function of phonon frequencies. The results for the ``opt.'' case are shown. 
The top and bottom panels are the SPS values for absorption and emission processes, respectively.}
\label{fig:sps}
\end{figure}

The three-phonon SPS of phonon $q$ is defined as
\begin{equation}
W^{\pm}_{q} = \frac{1}{N}{\sum_{q',q''}}
\left\{
    \begin{array}{c}
      n_{q''} - n_{q'} \\
      n_{q'} + n_{q''} + 1
    \end{array}
  \right\}
\delta(\omega_{q}-\omega_{q'}\pm \omega_{q''}) \delta_{\bm{q}\pm\bm{q}',\bm{q}''+\bm{G}}.
\end{equation}
When the SPS was calculated based on the SCP solution, the harmonic phonon frequencies were replaced by the temperature dependent anharmonic phonon frequencies, i.e., $\omega_{q}\rightarrow \Omega_{q}$. 
The calculated results are shown in Fig.~\ref{fig:sps}.
In the frequency range of 45--65 cm$^{-1}$, the SPS descreses by about 50\% due to the anharmonic renormalization of phonon frequencies.

\begin{figure}
\centering
\includegraphics[width=8.5cm,clip]{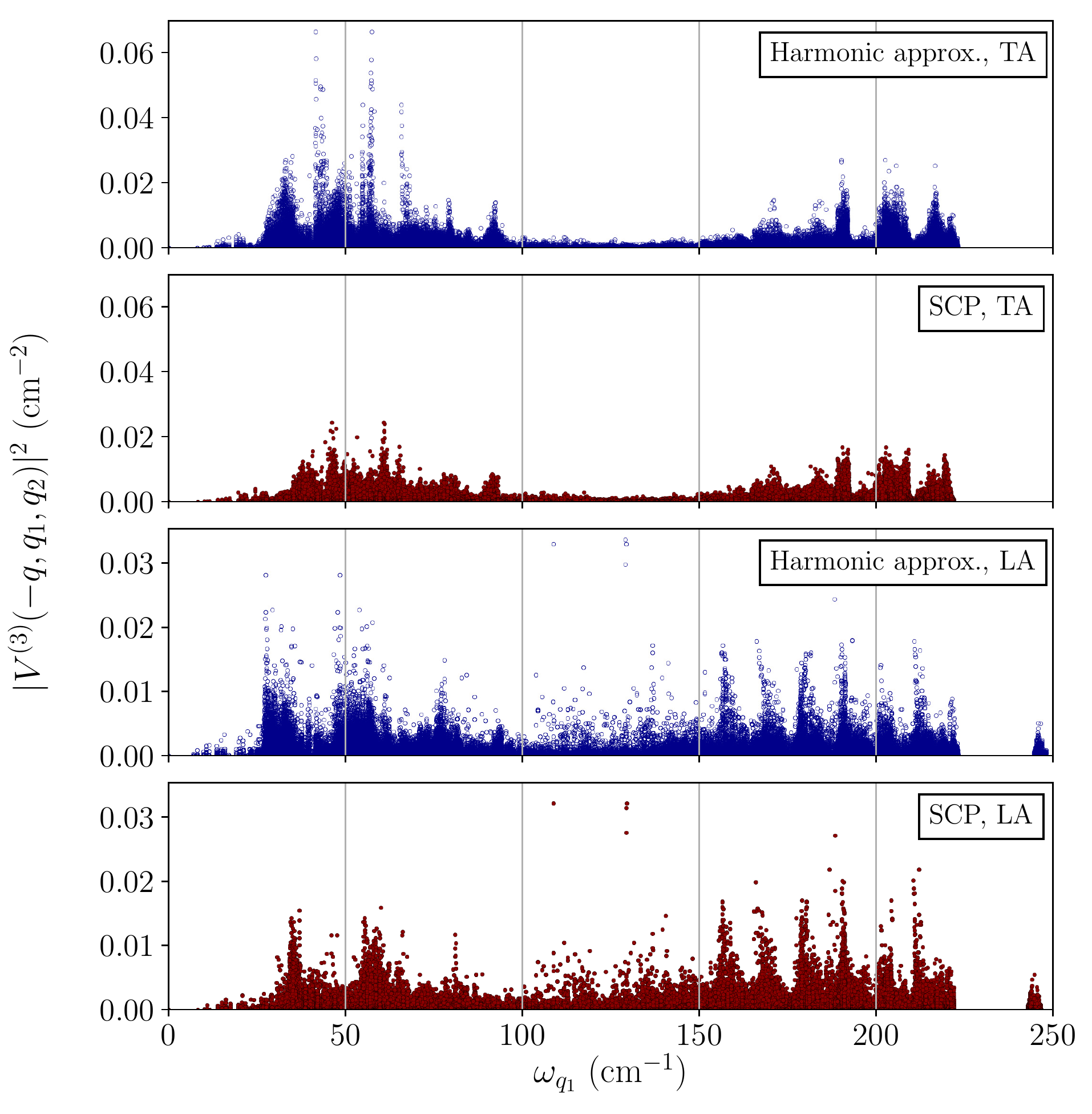}
\caption{The anharmonic coupling coefficients for acoustic modes at $\bm{q}=(\frac{1}{5}, 0, 0)$ as a function of $\omega_{q_{1}}$.}
\label{fig:V3}
\end{figure}

To highlight the importance of the change of anharmonic coupling coefficients, we also compare the components of $|V^{(3)}(-q;q_{1};q_{2})|^{2}=|\Phi(-q;q_{1};q_{2})|^{2}/8\omega_{q}\omega_{q_1}\omega_{q_2}$ calculated with and without the temperature renormalization of phonon frequencies.
The computational results are shown in Fig.~\ref{fig:V3}. 
Here, we selected transverse acoustic (TA) and longitudinal acoustic (LA) modes at $\bm{q} = (\frac{1}{5}, 0, 0)$.
The frequencies of these two acoustic modes are 13.8 and 22.1 cm$^{-1}$ within the harmonic approximation and 14.1 and 22.7 cm$^{-1}$ within the
self-consistent phonon theory. Also, the components of $|V^{(3)}(-q;q_{1};q_{2})|^{2}$ are shown only for the pairs $(q_{1},q_{2})$ satisfying the
momentum- and energy-conservation laws. Figure \ref{fig:V3} shows that the coupling coefficient $|V^{(3)}(-q;q_{1};q_{2})|^{2}$ of low-frequency acoustic modes descreases by the anharmonic renormalization of phonon frequencies and polarization vectors.

\clearpage

\end{document}